\documentclass[paper]{JHEP3}

\usepackage[dvips]{graphics}
\usepackage{rotating}
\usepackage{epsfig}
\usepackage{amsmath}
\def\lesssim{\mathrel{\hbox{\rlap{\hbox{\lower5pt\hbox{$\sim$}}}\hbox{$<$}}}}
\def\gtrsim{\mathrel{\hbox{\rlap{\hbox{\lower5pt\hbox{$\sim$}}}\hbox{$>$}}}}

\newcommand{\ntrl}[1]{\chi^0_#1}

\def\beq{\begin{equation}}   %
\def\eeq{\end{equation}}   %
\def\ben{\begin{eqnarray}}   %
\def\een{\end{eqnarray}}   %
\def\nn{\nonumber}

\title{Displaced Higgs production in type III seesaw}
\author{ Priyotosh Bandyopadhyay$^a$ $\mbox{ and } $
Eung Jin Chun$^b$\\
Korea Institute for Advanced Study,
 Hoegiro 87, Dongdaemun-gu,Seoul 130-722, Korea
\email{$^a$priyotosh@kias.re.kr, $^b$ejchun@kias.re.kr}
 }
\abstract{ We point out that the type III seesaw mechanism
introducing fermion  triplets predicts peculiar Higgs boson
signatures of displaced vertices with two b jets and  one or two
charged particles which can be cleanly identified. In a
supersymmetric theory, the scalar partner of the fermion triplet
contains a neutral dark matter candidate which is almost
degenerate with its charged components. A Higgs boson can be
produced together with such a dark matter triplet in the cascade
decay chain of a strongly produced squark or gluino. When the next
lightest supersymmetric particle (NLSP) is bino/wino-like, there
appears a Higgs boson associated with two charged tracks of a
charged lepton and a heavy charged scalar at a displacement larger
than about 1 mm.  The corresponding production cross-section is
about 0.5 fb for the squark/gluino mass of 1 TeV. In the case of
the stau NLSP, it decays mainly to a Higgs boson and a heavy
charged scalar whose decay length is larger than 0.1 mm for the
stau NLSP mixing with the left-handed stau smaller than 0.3. As
this process can have a large cascade production $\sim 2$ pb for
the squark/gluino mass $\sim 1$ TeV, one may be able to probe it
at the early stage of the LHC experiment. }

\preprint{KIAS-P10016}

\begin{document}

\section{Introduction}

The observed neutrino masses and mixing may arise from a TeV scale
seesaw mechanism. This implies that new particles responsible for
the seesaw mechanism have TeV-scale masses and their Yukawa
couplings to neutrinos, denoted by $y_\nu$, are as small as the
electron Yukawa coupling.  This comes from the fact that the
seesaw mass of neutrinos are given by $m_\nu \sim y_\nu^2 v^2/M$
where $v=174$ GeV is the Higgs vacuum expectation value and $M$ is
the new particle mass. One has $y_\nu \sim 10^{-6}$ for $m_\nu
\sim 0.01$ eV and $M=1$ TeV.  As a consequence of a small neutrino
Yukawa coupling, TeV-scale seesaw particles can leave displaced
vertices which are observable in the LHC detectors. Furthermore,
neutrino Yukawa couplings usually involve a Higgs boson and thus
it can be produced  at such displaced vertices.  Since this gives
a clean signature free from backgrounds, one may be able to probe
a light Higgs boson through its main decay channel $h \to
b\bar{b}$.

A typical example realizing such a feature is the type III seesaw
mechanism \cite{Foot88} where a massive $SU(2)_L$ triplet fermion
$\Sigma=(\Sigma^+, \Sigma^0, \Sigma^-)$ carrying no hypercharge
under $U(1)_Y$ is introduce to provide the neutrino Yukawa
coupling; $y_\nu L H_2 \Sigma$.  Rich phenomenology of the type
III seesaw has been studied  in
Refs.~\cite{Hambye08,Aguila08,Arhrib09,Bandyo09,Li09}. In this model, the
seesaw particles $\Sigma^{\pm,0}$ are produced through electroweak
interactions and there main decay modes are $\Sigma^{\pm}
(\Sigma^0) \to l^\pm (\nu) h$ through the Yukawa coupling $y_\nu$.
The corresponding decay length $\tau_\Sigma$ of $\Sigma^{\pm,0}$
can be long enough to be measured at the LHC if $y_\nu$ is small
enough. For instance, $\tau_\Sigma = 0.6$ mm for $y_\nu^2
v_2^2/M=$ meV and $M=500$ GeV.  The  production cross-section of
the process $\Sigma^{\pm} (\Sigma^0) \to l^\pm (\nu) b\bar{b}$ is
in the range of 1 pb $-$ 1 fb for the mass range $M=0.2 - 1$ TeV
\cite{Hambye08}. This can be compared to the cross-section
for an important Higgs discovery channel $h \to
\gamma\gamma$. This process suffers a huge background and the
expected cross-section for the Higgs boson plus two jet analysis
goes down to about 1 fb \cite{Atlas}. Thus, one may anticipate the
first Higgs discovery through the production of the triplet in the
type III seesaw mechanism.

\medskip

In this work, we consider a cascade production of displaced Higgs
bosons in the supersymmetric type III seesaw.  Our focus is on the
parameter region of large triplet fermion mass $M\sim 1$ TeV and
large neutrino Yukawa $y_\nu$ corresponding the atmospheric
neutrino mass scale $m_\nu \approx 0.05$ eV for which the triplet
fermion decays almost promptly  and thus leaves no observable
displace vertices.  An interesting aspect of the supersymmetric
type III seesaw is that the observational evidences for dark
matter and neutrino mass can be correlated in a way that the
supersymmetric partner $\tilde{\Sigma}$ of the triplet fermion
$\Sigma$ is the dark matter particle. More precisely, the neutral
component $\tilde{\Sigma}^0$ of the scalar triplet can be the
lightest supersymmetric particle (LSP) and thus can be dark matter
if R-parity is conserved.  In this case, next lightest
supersymmetric particles (NLSPs) are produced by cascade decays of
strongly produced squarks or gluinos, and  finally decay to a
Higgs boson associated with the dark metter $\tilde{\Sigma}^0$ or
its charged companion $\tilde{\Sigma}^\pm$.  The NLSP decay
involves additional suppression other than the neutrino Yukawa
coupling, and thus the decay length becomes much longer than the
$\Sigma$ decay discussed above. Then there will be observable
displaced vertices with two $b$ jets  and one or two charged
particles depending on what the NLSP is. For our discussion, we
will take a neutralino or a  stau as the NLSP. Recall that various
astrophysical and cosmological observations put a rather strong
bound on the dark matter mass: $m_{\tilde{\Sigma}^0} \gtrsim 520$
GeV \cite{Chun09}. This pushes up all the ordinary supersymmetric
particle masses and thus a compact supersymmetric spectrum is
needed to increase the cascade Higgs boson production rate.

\medskip

This paper is organized as follows. In Section 2, we will
introduce the type III seesaw mechanism realizing the dark matter
triplet and present all the vertices needed for our calculation.
Section 3 presents our main results on the cascade production of
the Higgs boson at displaced vertices taking the NLSP as a
neutralino or a stau. We conclude in Section 4.

\section{Type III seesaw and dark matter triplet}

\subsection{Supersymmetric type III seesaw}

The type III seesaw mechanism introduces real $SU(2)_L$ triplets
with $Y=0$.  Using the matrix representation, the triplet field
can be written as
 \ben {\bf \Sigma}=&\Sigma_i\cdot \sigma_i
=&\begin{pmatrix}
\Sigma^0 & \sqrt{2}\Sigma^+  \\
\sqrt{2}\Sigma^- &  -\Sigma^0
\end{pmatrix}
\een where $\Sigma^{\pm}={1\over\sqrt{2}}(\Sigma_1 \mp
i\Sigma_2)$. The gauge invariant superpotential is then
\begin{equation} \label{WIII0}
 W_{III} = y_\nu L^T i\sigma_2 {\bf\Sigma} H_2  + {1\over4} M \mbox{Tr}({\bf\Sigma}^2)\,,
\end{equation}
where we suppressed lepton flavor indices and the scalar component
of ${\bf\Sigma}$ is assumed to  form the dark matter triplet. We
will assume one flavor dominance for our calculations. Expanding
the above superpotential in components, we get
\begin{equation} \label{WIII}
 W_{III} = - y_\nu \left[\sqrt{2}\, \l\Sigma^+ H^0_2  + \, \nu \Sigma^0 H^0_2
 +\sqrt{2}\, \nu\Sigma^- H^+_2  +\, \l \Sigma^0 H^+_2 \right]
  +  M \Sigma^+\Sigma^- +{M\over2}\Sigma^0\Sigma^0\,.
\end{equation}
Integrating out the heavy triplet fields one obtains the seesaw neutrino mass matrix.
Here we define the effective neutrino mass associated with the dark matter triplet
\begin{equation} \label{mnu}
 \tilde{m}_\nu = {|y|^2 v_2^2 \over M}
 \,,
\end{equation}
where $v_2  = \langle H_2^0 \rangle$.

\medskip

Let us now briefly describe the properties of the scalar dark
matter triplet   $\tilde{\Sigma}$ \cite{Chun09}.  Including
supersymmetric and soft supersymmetry breaking terms, we get the
mass terms of the dark matter triplet from
\begin{eqnarray}\label{smass}
V_{mass}&=&(M^2  +  \tilde{m}^2) (|\tilde{\Sigma}^+|^2  + |\tilde{\Sigma}^+|^2)
+  (M^2  + \tilde{m}^2) |\tilde{\Sigma}^0|^2\,\\\nn
& +& BM \left[\tilde{\Sigma}^+\tilde{\Sigma}^- +  {1\over 2}\tilde{\Sigma}^0\tilde{\Sigma}^0
 +  h.c.\right] + {g^2\over 8} \left[|H_1^0|^2- |H_2^0|^2 +2 |\tilde{\Sigma}^+|^2
 -2 |\tilde{\Sigma}^-|^2\right]^2\,,
\end{eqnarray}
where we assume that $B$ is positive definite without loss of
generality.  Because of the off-diagonal $B$ term, the mass
eigenstates become
$\tilde{\Sigma}^0_{1,2}={1\over{\sqrt{2}i}}(\tilde{\Sigma}^0 -
\tilde{\Sigma}^{0*})$,
$\tilde{\Sigma}^0_{2}={1\over{\sqrt{2}}}(\tilde{\Sigma}^0  +
\tilde{\Sigma}^{0^*})$ and
$\tilde{\Sigma}^+_{1,2}={1\over{\sqrt{2}}}(\tilde{\Sigma}^+  \mp
\tilde{\Sigma}^{-*})$. The neutral scalar components with $T_3=0$
take the mass-squareds given by
\begin{equation} \label{mm0}
 m^2_{\tilde{\Sigma}^0_{2,1}} = M^2 +\tilde{m}^2 \pm B M \,,
 \end{equation}
where $\tilde{m}$ is the soft supersymmetry breaking mass.
Including the D-term contribution, the mass-squared eigenvalues of
the charged scalar components $\tilde{\Sigma}^\pm$ carrying
$T_3=\pm1$ are
\begin{equation} \label{mmpm}
 m^2_{\tilde{\Sigma}^\pm_{2,1}} = M^2 +\tilde{m}^2 \pm
 \sqrt{B^2 M^2 + c_W^4 m_Z^4 c^2_{2\beta} }\,.
 \end{equation}
where $c_W$ is the cosine of the weak mixing angle and the angle
$\beta$ is defined by $t_\beta= v_2/v_1$. The lighter states
$\tilde{\Sigma}_1$ have the  mass splitting $\Delta
m_{\tilde{\Sigma}} \equiv m_{\tilde{\Sigma}_1^\pm} -
m_{\tilde{\Sigma}_1^0}$ which is composed of the negative
tree-level contribution [Eqs.~(\ref{mm0}, \ref{mmpm})]  and the
positive one-loop contribution [$\Delta m_{\rm loop} \simeq 167$
MeV] \cite{Cirelli05}. Thus the total mass splitting $\Delta
m_{\tilde{\Sigma}}$ is smaller than 167 MeV but can remain
positive depending on the values of $BM$. In this case,
$\tilde{\Sigma}_1^0$ can be the LSP dark matter, and
$\tilde{\Sigma}_1^{\pm}$ decays to $\pi^\pm\tilde{\Sigma}_1^0$ or
$e^\pm\nu_e\tilde{\Sigma}_1^0$. The charged scalar triplets have
decay length larger than 5.5 cm and thus leaves slowly-moving and
highly-ionizing tracks that can be detected at the LHC
\cite{Chun09}.

\subsection{Couplings of dark matter triplet}

From the superpotential in Eq.~(\ref{WIII}) one can find the
F-term couplings involving the neutral Higgs bosons $H_{1,2}^0$:
\begin{equation}
V_F = y_\nu \left[ -\sqrt{2} M\, \tilde{\l} \tilde{\Sigma}^{-*}
{H^0_2}
   -  M\, \tilde{\nu}\tilde{\Sigma}^{0*} {H^0_2}
   +\sqrt{2} \mu\, \tilde{\l} \tilde{\Sigma}^{+} {H^0_1}^*
   +  \mu \,\tilde{\nu}\tilde{\Sigma}^{0} {H^0_1}^*\right] + h.c.
   \,,
\end{equation}
where the $\mu$ terms come from the Higgs bilinear term in the
superpotential, $W_H = \mu H_1 H_2$. The soft supersymmetry
breaking interactions are given by
\begin{equation}
V_{\rm soft}= -\sqrt{2}y_\nu A\, \tilde{\l}\tilde{\Sigma}^+ H^0_2
              - y_\nu A\, \tilde{\nu} \tilde{\Sigma}^0 H^0_2 +h.c. \,,
\end{equation}
where $A$ is the trilinear soft parameter.   Now, taking only the
vertices of $\tilde{\Sigma}^0_1$ and $\tilde{\Sigma}^\pm_1$,  we
have
 \begin{eqnarray} \label{Vsigma}
  -{\cal
L}_{\rm scalar}
 &=& y_\nu (M-A) \left[ \tilde{\l}\tilde{\Sigma}^+_1 H^0_2
 -{1\over2}\tilde{\nu}_I \tilde{\Sigma}^0_1 H^0_2 \right] \\
 \nonumber
&& + \, y_\nu \mu \left[ \tilde{\l} \tilde{\Sigma}_1^{+} {H^0_1}
 - {1\over2} \tilde{\nu}_I\tilde{\Sigma}_1^{0} {H^0_1} \right]
+h.c.\,.
 \end{eqnarray}
  where $\tilde{\nu}_I\equiv \sqrt2 \Im(\tilde{\nu})$.
 This gives rise to the $\tilde{l}^\pm$--$\tilde{\Sigma}_1^\pm$
and $\tilde{\nu}$--$\tilde{\Sigma}_1^0$ mixing and the
corresponding mixing angles are given by
\begin{eqnarray} \label{thetaB}
\theta_{\tilde{l}}&\approx&{y_\nu v_2
(M-A+\mu/\tan\beta)\over{(m_{\tilde{\l}}^2
- m^2_{\tilde{\Sigma^{+}_1}})}} \quad\mbox{and}\\
 \label{thetaB2}
\theta_{\tilde{\nu}}&\approx&-{y_\nu v_2
(M-A+\mu/\tan\beta)\over{(m_{\tilde{\nu}}^2 -
m^2_{\tilde{\Sigma^{0}_1}})}}
\end{eqnarray}
for the charged and neutral parts, respectively.\footnote{Note
that these mixing angles contain the $\mu$ term which was ignored
in Ref.~\cite{Chun09}.} Taking only  the real degrees of freedom
of $H^0_{1,2}$: $H^0_{1}=v_{1}-\sin\alpha\, h/\sqrt2$ and
$H^0_2=v_2+\cos\alpha\,h/\sqrt2$, and neglecting heavy Higgs
bosons in Eq.~(\ref{Vsigma}), we get the light Higgs boson
couplings:
\begin{eqnarray}
\label{vrt3} -{\cal L}_{h} & =& {y_\nu \cos\alpha\over\sqrt2}
(M-A-\mu\tan\alpha)\,  h \left[\tilde{\l}^-\tilde{\Sigma}^+_1
+\tilde{\l}^+\tilde{\Sigma}^-_1 -\tilde{\nu}_I
\tilde{\Sigma}^0_1\right] \een

\begin{figure}[t]
\begin{center}
%
{\epsfig{file=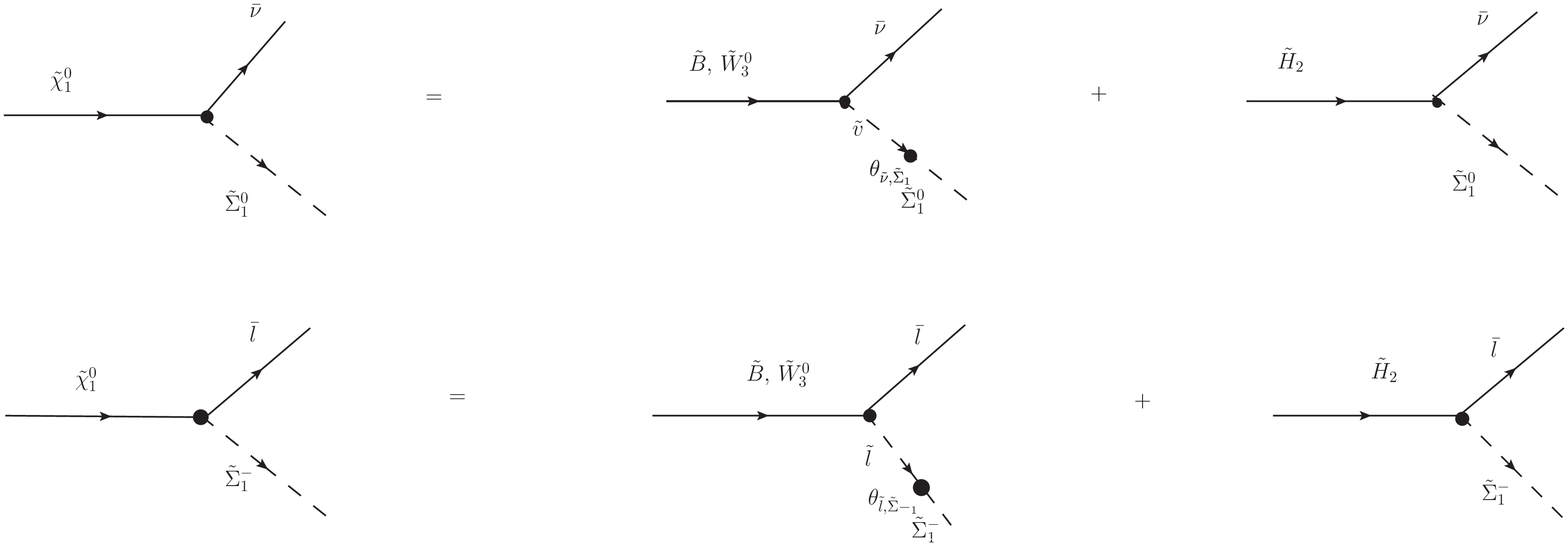,width=11.0 cm,height=7.0cm,angle=0}}
\caption{Feynman diagrams for the gaugino NLSP two-body decay:
$\ntrl1 \to \nu  \tilde{\Sigma}^{0}_1$ and $ l^\pm
\tilde{\Sigma}^\mp_1$.  }
\end{center}
\end{figure}
\begin{figure}[t]
\begin{center}
%
{\epsfig{file=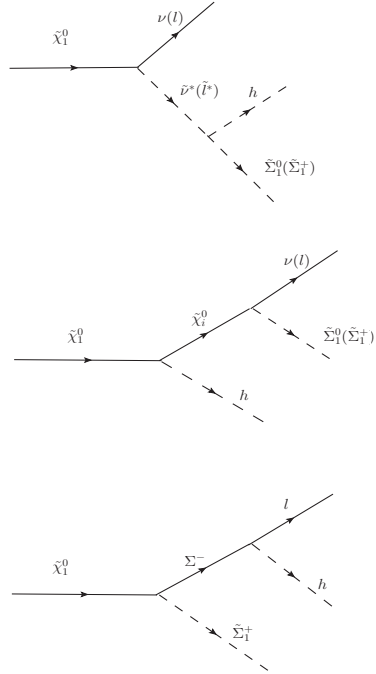,width=5.0 cm,height=9.0cm,angle=0}}
\caption{Feynman diagrams for the gaugino NLSP three-body decay:
$\ntrl1 \to  \nu h  \tilde{\Sigma}^{0}_1$ and $ l^\pm h
\tilde{\Sigma}^\mp_1$.}

\end{center}
\end{figure}

Note also that the fermion couplings from Eq.~(\ref{WIII}) give
the mixing between $l$ ($\nu$) and $\Sigma^-$ ($\Sigma^0$):
\begin{equation} \label{thetaF}
 \theta_l \approx {\sqrt2 y_\nu v_2 \over M}\quad\mbox{and}\quad
 \theta_\nu \approx {y_\nu v_2 \over M}\,.
\end{equation}
The mixing in Eqs.~(\ref{thetaB},\ref{thetaB2}, \ref{thetaF})
induces the gaugino--lepton--dark matter triplet interactions as
follows:
\begin{eqnarray}
 {\cal L}_{\rm gaugino} &=&
 -{g'\over2}  \left[ i \theta_{\tilde{\nu}}\, \tilde{B}\nu\tilde{\Sigma}^0_1
 + \theta_{\tilde{l}}\, \tilde{B} l \tilde{\Sigma}^+_1\right] \\\nn
&& +{g\over2} \left[ i \theta_{\tilde{\nu}}\,\tilde{W}_3  \nu\tilde{\Sigma}^0_1
     -(\theta_{\tilde{l}} + 2\theta_l) \,\tilde{W}_3  l \tilde{\Sigma}^+_1 \right] + h.c.\,.
\end{eqnarray}

Together with the fermion-fermion-scalar couplings from Eq.~(2.3),
Eqs.~(2.13,2.15) define the interactions required for our
calculation.  Figs.~1 and 2 summarize the neutralino NLSP decay
diagrams relevant for our analysis.

\section{Cascade Higgs production at displaced vertices}

\subsection{Bino-like NLSP}
Let us first consider the bino-like NLSP as a typical example.
Since our dark matter candidate has a heavy mass
$m_{\tilde{\Sigma}}>520$ GeV, we choose a compact supersymmetric
spectrum to enhance a cascade production of the light higgs boson
and dark matter triplet. For the illustration of our main points,
we take input parameters as follows:

$m_{\tilde{\Sigma}}=550$ GeV , \quad M=1 TeV

$m_{\tilde{q},\, \tilde{g}}=900 \, \rm{GeV}, \quad \quad m_{\tilde{\l}, \,
\tilde{\nu}}=900 \, \rm{GeV} $

$m_{A}=600 \, \rm{GeV}, \quad \mu=-2000 \, \rm{GeV}, \quad \tan{\beta}=10\quad
A_t=-1000\, \rm{GeV}, \quad A_{b,\tau}=0$

$M_1=750 \, \rm{GeV}, \quad M_2=800 \, \rm{GeV}, \quad M_3=900\, \rm{GeV}\,.$\\
With this set of input parameters, we get the corresponding Higgs
mass spectrum and the gaugino mass spectrum as follows.

$m_{h}=119 \, \rm{GeV}, \quad m_H=599 \, \rm{GeV}, \quad
m_A=600\,\rm{GeV},\quad m_{H^{\pm}}=604\, \rm{GeV},$

$m_{\ntrl1}=745 \, \rm{GeV}, \quad m_{\ntrl2}=810 \, \rm{GeV}, \quad
m_{\ntrl3}=1983\,\rm{GeV}, \quad m_{\ntrl4}=1984\, \rm{GeV}.$\\
%
%
\begin{figure}[t]
\begin{center}
%
{\epsfig{file=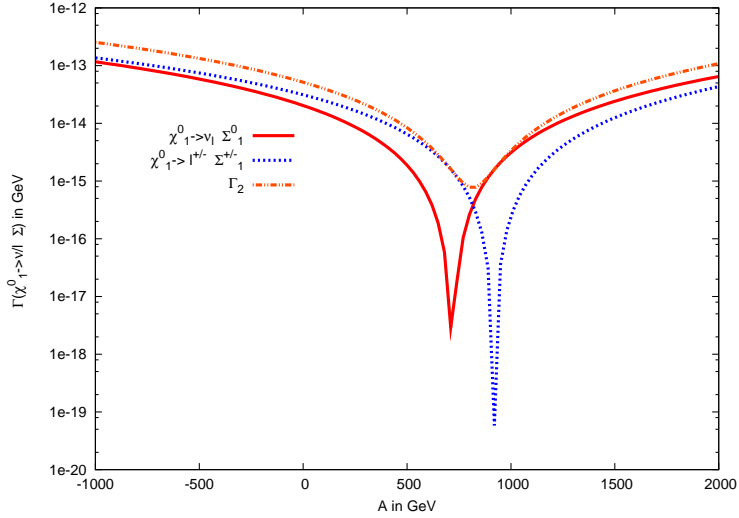,width=7.0 cm,height=10.0cm,angle=-90}}
\caption{Partial and total two-body decay width  of $\ntrl1\to
\nu_e(\l) \Sigma^0_1(\Sigma^+_1)$ with M=1 TeV and A varied in the
x axis. The effective neutrino mass is taken to be $\tilde{m}_\nu
=0.05$ eV. }
\end{center}
\end{figure}
\begin{figure}[t]
\begin{center}
%
{\epsfig{file=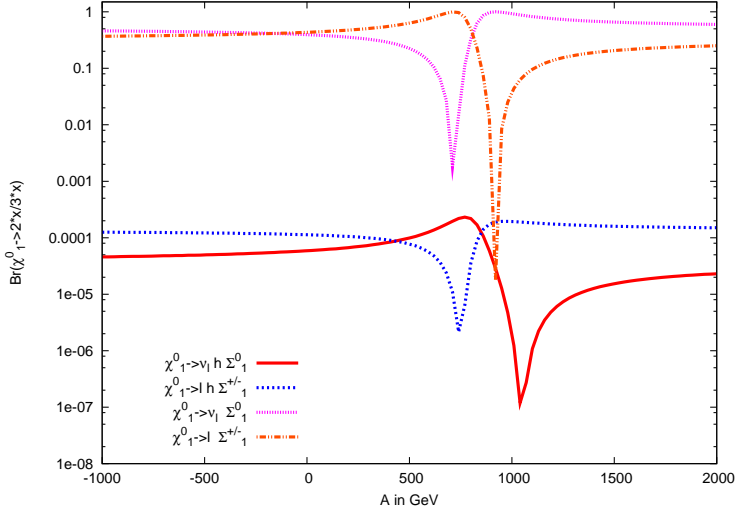,width=7.0 cm,height=10.0cm,angle=-90}}
\caption{Branching fractions of different modes on  of $\ntrl1$.
Input paramters are the same as in Fig.~3.}
\end{center}
\end{figure}
\begin{figure}[t]
\begin{center}
%
{\epsfig{file=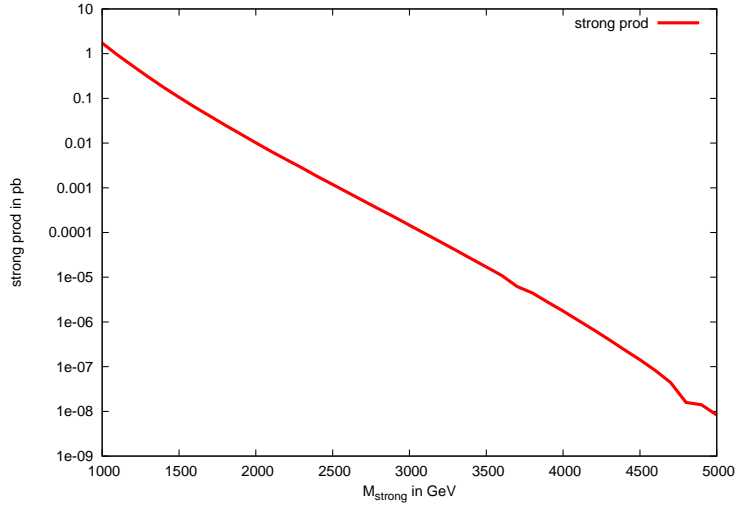,width=7.0 cm,height=10.0cm,angle=-90}}
\caption{Strong Supersymmetric production cross-section in pb in
y-axis and $M_{\tilde{q}}=M_{\tilde{g}}=M_{\rm strong}$ in GeV in
x-axis}
\end{center}
\end{figure}
\begin{figure}[t]
\begin{center}
%
{\epsfig{file=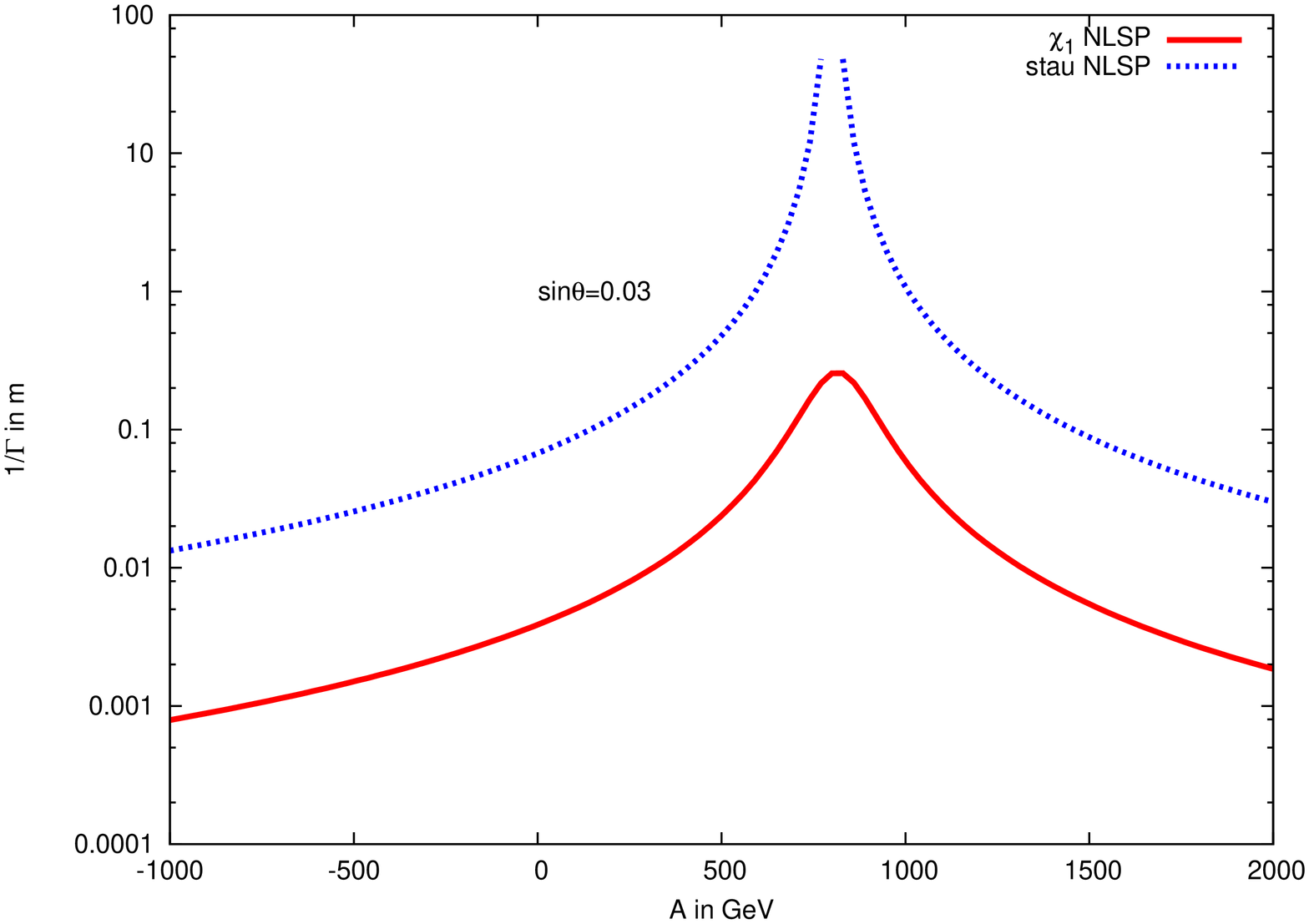,width=7.0 cm,height=10.0cm,angle=-90}}
\caption{Decay length in meter for the bino and stau NLSP  with
M=1 TeV and $\tilde{m}_\nu=0.05$ eV. Both decay lengths are
inversely proportional  to $\tilde{m}_\nu$ and the stau decay
length scales with $1/\sin^2\theta$.  For this plot we take
$\sin\theta=0.03$ and $m_{\tilde{\tau}_1}=700$ GeV }
\end{center}
\end{figure}
%

\noindent
 Now we calculate the two- and three-body decays of the
bino-like NLSP, $\ntrl1 \to \nu  \tilde{\Sigma}^{0}_1, \; l^\pm
\tilde{\Sigma}^\mp_1$ and $\ntrl1 \to \nu  h
\tilde{\Sigma}^{0}_1,\; l^\pm h \tilde{\Sigma}^\mp_1$,  whose
Feynman diagrams are shown in Figs.~1 and 2.  For two-body decay widths
there are two contributions; one through the mixing with sleptons and the 
other one is through the direct coupling to $H_2$ which is evident from Fig.~1.
 In Fig.~3, we plot
the two-body decay rates varying the $A$ parameter of the neutrino
Yukawa couplings.  The deeps in the plot appear due to the
cancellation of different contributions shown in Fig.~1. In particular
the cancellation for the charged leptons and the neutrinos occur at different 
points owing to the difference of $\sqrt{2}$ in the field definitions  \ref{WIII0}. Due to this the total two-body decay is not much suppresed as the partial
decay widths. From Fig.~2 we can see the different Feynman diagrams contributing to the three-body decays. In the case of charged lepton three-body decay mode there is one more contribution through the off-shell $\Sigma^{\pm}$.
 Note that
the decay rates are proportional to the effective neutrino mass
$\tilde{m}_\nu$ (\ref{mnu})  which is taken to be $0.05$ GeV
corresponding the atmospheric neutrino mass scale.  In Fig.~4, we
present the branching ratios of the two-body decays as well as
three-body decays involving the Higgs production. The Higgs
production is suppressed due to a  small three-body phase space.
However, the cascade Higg boson production can be sizable if the
supersymmetric particles have a compact mass spectrum near 1 TeV.
To show this, we calculate the gluino/squark production
cross-section as a function of a common gluino/squark mass in
Fig.~5. The total strong supersymmetric production cross-section
for our mass spectrum, $m_{\tilde{q}}\approx m_{\tilde{g}}\approx
900$ GeV, is 2.65 pb. Thus, we get the cross-section of about 0.5
fb for the Higgs production accompanied with two charged
particles, $\chi^0_1 \to h l^\pm \tilde{\Sigma}^\mp$, as its
branching fraction is about $2\times10^{-4}$.

Note that the displacement of the $\chi^0_1$ decay is lager than 1
mm  for $\tilde{m}_\nu=0.05$ eV as shown in Fig.~6. The effective
neutrino mass of 0.05 eV corresponds to the atmospheric neutrino
oscillation scale  is the maximum value excluding the degenerate
neutrino mass pattern. Thus, the displacement is generally
expected to be larger than  1 mm as it is inversely proportional
to $\tilde{m}_\nu$. This  can  be compared with the displacement
of the fermion triplet $\Sigma$ decay \cite{Hambye08} which
becomes $\tau_\Sigma \approx 3 \mu\mbox{m}$ for the same
parameters as above: $\tilde{m}_\nu = 0.05$ eV and $M=1$ TeV. The
bino-like NLSP decay length is larger than the  $\Sigma$ decay
because the former involves an additional gauge coupling $g'$ and
the mixing angle $\theta_{\tilde{l}, \tilde{\nu}}$ typically
smaller than the neutrino Yukawa coupling $y_\nu$.  Furthermore,
the Higgsino contribution to the decay [see Fig.~1] is generically
suppressed by its small mixture in $\chi^0_1$.

\medskip

 LHC detectors should be able to reconstruct the position
of a displaced vertex from the $\chi^0_1$ decay by observing two
charged tracks induced by a lepton $l^\pm$ and a dark matter
triplet component $\tilde{\Sigma}^\mp$. Depending on the mass gap
$\Delta m_{\tilde{\Sigma}}$,  $\tilde{\Sigma}^\pm$ leaves a
slowly-moving and highly-ionizing track of 5.5 cm -- 6.3 m
 \cite{Chun09} before it decays to $\pi^\pm \tilde{\Sigma}^0$.
 Such signals can be probed at the early stage of the LHC
 experiment if the squark/gluino  is not too heavy so that
 the NLSP production cross-section is large enough [see Fig.~5].
This is a unique signature of the dark matter triplet of type III
seesaw.  In addition to this, the Higgs bosons associated with two
charged particles are produced to yield displaced vertices with
$l^\pm \tilde{\Sigma}^\mp b \bar{b}$.  These are clean signatures
free from backgrounds and may help us to probe the Higgs boson
property confirming the major Higgs decay to $b\bar{b}$ although
its cross-section is small. Here let us compare the above events
with the electroweak production of the fermion triplet components
$\Sigma^{\pm,0}$ and their decays.  The cross-section of the
$\Sigma^{\pm} \Sigma^{\mp,0}$ is about 2 fb for our parameter
region \cite{Hambye08} whereas the cascade production of $pp \to
\chi_1^0 \chi_1^0 \to h l^+ \tilde{\Sigma}^- \, l^\pm
\tilde{\Sigma}^\mp$ is about 0.5 fb. Note that we can have
same-sign  and opposite-sign dileptons with the same possibility.
The bino-like NLSP decay leaves sizable displaced vertices whereas
$\Sigma^{\pm,0}$ promptly decays $l^\pm h$ or $\nu h$ as discussed
above.  Recall that the usual Higgs signals of $h\to \gamma\gamma$
becomes about 1 fb in two jet analysis \cite{Atlas}. Thus the
cascade production of displaced Higgs bosons associated with two
charged tracks can be useful to probe the Higgs property through
its main decay $h\to b\bar{b}$.

\subsection{Wino-like NLSP}

The wino-like NLSP have similar properties as the bino-like NLSP
except that its decay length is generically shorter. This comes
from the larger gauge coupling constant ($g> g'$) which leads to a
suppression factor of $(g'/g)^2 \sim 0.5$ for the decay length.

\subsection{Higgsino-like NLSP}

In the case of the Higgsino-like NLSP, the decay length becomes
even shorter.  Since it decays to $l^\pm \tilde{\Sigma}^\mp$ or
$\nu \tilde{\Sigma}^0$ through the neutrino Yukawa coupling
$y_\nu$, its decay length is comparable to that of $\Sigma$ and it
cannot lead to a displaced Higgs production.

\subsection{Stau NLSP}

In the case of the stau (or any slepton) NLSP, the two-body decay
through the neutrino Yukawa coupling $y_\nu$, $\tilde{\tau}_1^\pm
\to h \tilde{\Sigma}^\pm $, is the main decay mode.  Thus, it
leads to a huge cascade production cross-section (2.65 pb) for our
parameter choice. Even for a larger squark/gluino mass, one may
have a sizable production. For instance, requiring the
cross-section of 1 fb, th signals of $\tilde{\Sigma}^\pm b\bar{b}$
can be observed for the  squark/gluino mass of 2.5 TeV as shown in
Fig.~5.
 Whether it
can have a observable displacement depends on the mixing angle of
the stau NLSP to the left-handed stau.
 The dotted (blue) line in Fig.~6 shows the displacement which is inversely
proportional to the stau NLSP mass and the mixing angle.  One can
see that the decay length can be larger 100 $\mu$m as long as the
mixing parameter $\sin\theta$ is smaller than 0.3.

\section{Conclusion}

It is pointed out that the type III seesaw model for generating
light neutrino masses can lead to a displaced Higgs production
which can play an important role for the Higgs discovery at the
LHC.  Higgs bosons can be produced through the decay of a heavy
seesaw particle and a sizable displacement  occurs primarily due
to a small neutrino Yukawa coupling.  A peculiar feature of the
type III seesaw, introducing a $SU(2)_L$ triplet seesaw particle,
is  that the charged component yields a Higgs production
associated with a charged particle. This makes observable a light
Higgs boson through its main decay to two bottom quarks.

Considering the supersymmetric type III seesaw mechanism where the
neutral  scalar component $\tilde{\Sigma}^0$ of the triplet
superfield can form dark matter, we investigated  the Higgs
production through the cascade decay of strongly produced squarks
and gluinos assuming a neutralino and a stau NLSP.  In the case of
the neutralino (bino or wino) NLSP, Higgs bosons are produced by
three body decays, $\chi^0_1 \to h l^\pm \tilde{\Sigma}^\mp$ or $h
\nu \tilde{\Sigma}^0$. Although this process is suppressed
compared to the main two body decay, $\chi^0_1 \to l^\pm
\tilde{\Sigma}^\mp$ or $\nu \tilde{\Sigma}^0$, the cross-section
of the cascade Higgs production is larger than about 0.5 fb for
the squark/gluino mass below 1 TeV.  This requires a rather
compact supersymmetric spectrum as the dark matter mass has a
tight lower bound of 520 GeV coming from various astrophysical and
cosmological observations. Once such a spectrum is realized,  the
cascade Higgs signals can be cleanly identified as it involves two
charged tracks corresponding to a charge lepton and a heavy scalar
charged particle $\tilde{\Sigma}^\pm$ which leaves slowly-moving
and highly-ionizing tracks of the length longer than 5.5 cm.
Furthermore, the decay length of $\chi^0_1$ turns out to be
typically larger than 1 mm even for the largest neutrino Yukawa
coupling corresponding to the atmospheric neutrino mass scale
$m_\nu=0.05$ eV. Thus we would be able to find the Higgs boson
through the channel $h\to b{\bar{b}}$ in addition to a
conventional channel of $h\to \gamma\gamma$.

In the case of stau NLSP, the cascade production of Higgs bosons
is much more efficient as Higgs bosons arise through the main
decay channel $\tilde{\tau}_1^\pm \to h \tilde{\Sigma}^\pm$. The
stau decay length turns out to be larger than 0.1 mm if
$\sin\theta \lesssim 0.3$ where $\theta$ is the stau NLSP mixing
angle with the left-handed stau. Then, the displaced Higgs signal
can be observed even at the earlier stage of the LHC experiment as
the cascade production cross-section lies in the range of 2 pb --
1 fb for the squark/gluino mass of 1 TeV -- 2.5 TeV.

\bigskip

{\bf Acknowledgments:} E.J.C. was supported by Korea Neutrino
Research Center through National Research Foundation of Korea
Grant (2009-0083526).

\end{document}